\pdfoutput=1

\documentclass[twocolumn]{aastex631}
\usepackage{amsmath}
\usepackage{graphicx}
\usepackage{mathrsfs}
\usepackage{color}
\usepackage{url}
\usepackage{CJKutf8}
\usepackage{ulem}
\usepackage{wasysym}
\usepackage{multirow}
\usepackage{booktabs}

\received{2023}
\revised{\today}
\accepted{2024}

\shorttitle{C/2019 E3}
\shortauthors{Hui et al. 2023}

\begin{document}

\title{
Serendipitous Archival Observations of A New Ultra-distant Comet C/2019 E3 (ATLAS)
}

\correspondingauthor{Man-To Hui}
\email{mthui@must.edu.mo, manto@hawaii.edu}

\author{\begin{CJK}{UTF8}{bsmi}Man-To Hui (許文韜)\end{CJK}}
\affiliation{State Key Laboratory of Lunar and Planetary Science, 
Macau University of Science and Technology, 
Avenida Wai Long, Taipa, Macau}

\author{Robert Weryk}
\affiliation{Physics and Astronomy, The University of Western Ontario,
1151 Richmond Street,
London ON N6A 3K7, Canada}

\author{Marco Micheli}
\affiliation{ESA NEO Coordination Centre, Planetary Defence Office, Largo Galileo Galilei 1, I-00044 Frascati (RM), Italy}

\author{\begin{CJK}{UTF8}{bsmi}Zhong Huang (黃中)\end{CJK}}
\affiliation{State Key Laboratory of Lunar and Planetary Science, 
Macau University of Science and Technology, 
Avenida Wai Long, Taipa, Macau}

\author{Richard Wainscoat}
\affiliation{Institute for Astronomy, University of Hawai`i, 
2680 Woodlawn Drive, Honolulu, HI 96822, USA}


\begin{abstract}

We identified a new ultra-distant comet C/2019 E3 (ATLAS) exhibiting preperihelion cometary activity at heliocentric distances $\ga\!20$ au, making it the fourth member of this population after C/2010 U3 (Boattini), C/2014 UN$_{271}$ (Bernardinelli-Bernstein), and C/2017 K2 (PANSTARRS). From serendipitous archival data, we conducted analyses of the comet, finding that the activity was consistent with steady-state behaviour, suggestive of sublimation of supervolatiles, that the cross-section of dust increased gradually on the inbound leg of the orbit, varying with heliocentric distances as $r_{\rm H}^{-1.5 \pm 0.4}$, and that the dust was produced at a rate of $\ga\!10^2$ kg s$^{-1}$ within the observed timespan. Our modelling of the largely symmetric morphology of the comet suggests that the dust environment was likely dominated by mm-scale dust grains ejected at speeds $\la\!0.4$ m s$^{-1}$ from the sunlit hemisphere of the nucleus. Assuming a typical geometric albedo of 0.05 and adopting several simplistic thermophysical models, we estimated the nucleus to be at least $\sim\!3$ km across. We also measured the colour of the comet to be consistent with other long-period comets, except being slightly bluer in $g-r$. With our astrometric measurements, we determined an improved orbit of the comet, based upon which we derived that the comet is dynamically new and that its perihelion distance will further shrink due to the Galactic tide. We conclude the paper by comparing the known characteristics of the known ultra-distant comets.

\end{abstract}


\section{Introduction}
\label{sec_intro}

Recent years witnessed a burst of interest in research on ultra-distant comets, thanks to discoveries of three such objects, C/2017 K2 (PANSTARRS), C/2010 U3 (Boattini), and C/2014 UN$_{271}$ (Bernardinelli-Bernstein), which exhibited activity at unprecedentedly seen preperihelion heliocentric distances $r_{\rm H} \ga 20$ au \citep[][]{2017ApJ...847L..19J, 2017ApJ...849L...8M, 2018AJ....155...25H, 2019AJ....157..162H, 2021ApJ...921L..37B}. Deeply frozen for the majority of their lifetime, these comets are conceived to be the most primitive small bodies in the solar system bearing significant scientific importance. Greatly limited by the number of available samples, very little is known about how these comets are active at such large distances from the Sun. The current overall understanding is that cometary activity at $r_{\rm H} \ga 20$ au preperihelion is most likely driven by sublimation of supervolatiles such as CO and CO$_{2}$, which are reported to be abundant in comets \citep[e.g.,][]{2012ApJ...758...29A}. At such great distances from the Sun, the equilibrium surface temperature is $\la\!60$ K, which may be even too low for crystallisation of amorphous water ice \citep[e.g.,][]{2012AJ....144...97G}. Interestingly, some recent models predicted that even comets in the Oort cloud have been intensively processed by cosmic-ray bombardment, thereby depleting CO (but not CO$_{2}$ or CH$_{4}$) in the outermost $\sim\!10$ m of these cometary nuclei \citep{2020ApJ...890...89G,2020ApJ...901..136M}. This appears to be supported by the measured production rates of CO and CO$_{2}$ of dynamically new comets, whose activity tends to be dominated by CO$_{2}$ rather than CO \citep{2022PSJ.....3..247H}. In addition, observations of active ultra-distant comets pose a serious challenge to the classical comet model, which predicts no activity whatsoever for comets at these heliocentric distances, because drag forces from sublimating supervolatiles are not supposed to be strong enough to overcome interparticle cohesion \citep{2019AJ....157...65J}. In order to understand better how ultra-distant comets as well as comets in general are active, we feel the necessity to search for more comets of this kind to increase the available sample size. 

In this paper, we report on the fourth ultra-distant comet that exhibited preperihelion cometary activity at $r_{\rm H} \ga 20$ au, C/2019 E3 (ATLAS), using serendipitous prediscovery archival data. As the name suggests, the long-period comet was discovered by the Asteroid Terrestrial-impact Last Alert System (ATLAS) survey, on UTC 2019 March 5 at $r_{\rm H} = 13.6$ au \citep{2019MPEC....F...54Y}. The latest orbital solution by JPL Horizons shows that the current osculating heliocentric orbit of the comet is slightly hyperbolic (eccentricity $e = 1.002$) and highly inclined to the ecliptic plane (orbital inclination $i = 84\fdg3$), and that it recently reached its perihelion at $q = 10.3$ au in 2023 mid November.\footnote{See the solution by JPL Horizons at \url{https://ssd.jpl.nasa.gov/tools/sbdb_lookup.html\#/?sstr=2019e3}.} In the paper, we first detail the serendipitous archival observations of the comet in Section \ref{sec_obs}, present results in Section \ref{sec_rslt}, then discuss our analyses in Section \ref{sec_disc}, and finally summarise in Section \ref{sec_sum}.

\section{Observations}
\label{sec_obs}

\begin{deluxetable*}{cccccccccccC}
\tabletypesize{\footnotesize}
\tablecaption{Archival Serendipitous Observations and Viewing Geometry of Comet C/2019 E3 (ATLAS)
\label{tab:vgeo}}
\tablewidth{0pt}
\tablehead{
\colhead{Date} & \multicolumn{4}{c}{Archival Observations} &
\multicolumn{7}{c}{Viewing Geometry} \\ \cmidrule(lr){2-5} \cmidrule(lr){6-12}
\colhead{(UTC)} & Facility & Filter & \# images & Exposure (s) & $r_{\rm H}$ (au)\tablenotemark{a} & ${\Delta}$ (au)\tablenotemark{b} & $\alpha$ (\degr)\tablenotemark{c} & $\varepsilon$ (\degr)\tablenotemark{d} & $\theta_{-\odot}$ (\degr)\tablenotemark{e} & $\theta_{-{\bf V}}$ (\degr)\tablenotemark{f} & $\psi$ (\degr)\tablenotemark{g}
}
\startdata
2012 Mar 19 & PS1 & $w$ & 1 & 45 & 23.009 & 22.037 & 0.5 & 167.2 & 90.5 & 17.8 & -0.5 \\ \hline
2012 Nov 25 & PS1 & $w$ & 2 & 45 & 22.062 & 22.324 & 2.5 & 73.3 & 291.0 & 14.5 & +2.4 \\ \hline
2013 Feb 14 & PS1 & $r$ & 2 & 40 & 21.756 & 20.853 & 1.1 & 155.5 & 311.1 & 16.1 & +1.0 \\ \hline
2013 Feb 15 & PS1 & $w$ & 3 & 45 & 21.752 & 20.843 & 1.0 & 156.5 & 312.0 & 16.1 & +0.9 \\ \hline
2013 Mar 5 & PS1 & $w$ & 4 & 45 & 21.685 & 20.704 & 0.4 & 171.1 & 1.1 & 17.0 & +0.1 \\ \hline
2013 Apr 1 & DECam & $z$ & 1 & 100 & 21.584 & 20.680 & 1.2 & 154.2 & 94.9 & 18.4 & -1.1 \\ \hline
2013 Apr 3 & DECam & $z$ & 2 & 100 & 21.576 & 20.687 & 1.2 & 152.2 & 96.4 & 18.5 & -1.2 \\ \hline
2013 Apr 3 & PS1 & $w$ & 4 & 45 & 21.575 & 20.687 & 1.2 & 152.0 & 96.5 & 18.5 & -1.2 \\ \hline
2013 Apr 4 & OmegaCAM & $g$ & 2 & 50 & 21.573 & 20.690 & 1.3 & 151.3 & 97.0 & 18.5 & -1.3 \\ \hline
2013 Apr 7 & OmegaCAM & $r$ & 2 & 45 & 21.561 & 20.703 & 1.4 & 148.3 & 98.8 & 18.6 & -1.4 \\ \hline
2014 Feb 20 & DECam & $i$ & 3 & 160 & 20.354 & 19.431 & 1.0 & 158.5 & 325.7 & 16.3 & +0.8 \\
& & $g$ & 3 & 160 & & & & & & & \\ \hline
2014 Feb 20 & PS1 & $i$ & 1 & 45 & 20.353 & 19.428 & 1.0 & 158.8 & 326.3 & 16.3 & +0.8 \\ \hline
2014 Feb 28 & PS1 & $w$ & 4 & 45 & 20.323 & 19.365 & 0.7 & 164.8 & 345.4 & 16.8 & +0.4 \\ \hline
2014 Mar 11 & DECam & $i$ & 4 & 30 & 20.283 & 19.311 & 0.6 & 167.6 & 32.0 & 17.4 & -0.2 \\
& & $g$ & 2 & 30 & & & & & & & \\ \hline
2014 Dec 10 & PS1 & $i$ & 2 & 45 & 19.244 & 19.276 & 2.9 & 86.7 & 292.7 & 14.1 & +2.9 \\ \hline
2014 Dec 18 & PS1 & $i$ & 2 & 45 & 19.214 & 19.111 & 2.9 & 94.5 & 294.8 & 14.2 & +2.9 \\ \hline
2015 Jan 2 & PS1 & $i$ & 1 & 45 & 19.158 & 18.810 & 2.8 & 109.3 & 299.1 & 14.4 & +2.7 \\ \hline
2015 Jan 16 & PS1 & $r$ & 2 & 45 & 19.105 & 18.548 & 2.5 & 123.2 & 304.2 & 14.8 & +2.4 \\ \hline
2015 Jan 18 & PS1 & $w$ & 4 & 45 & 19.097 & 18.513 & 2.4 & 125.2 & 305.0 & 14.9 & +2.3 \\ \hline
2015 Jan 20 & PS1 & $w$ & 4 & 45 & 19.090 & 18.481 & 2.4 & 127.1 & 305.9 & 14.9 & +2.2 \\ \hline
2015 Jan 22 & PS1 & $w$ & 2 & 45 & 19.082 & 18.447 & 2.3 & 129.0 & 306.9 & 15.0 & +2.2 \\ \hline
2015 Mar 22 & PS1 & $w$ & 4 & 45 & 18.860 & 17.928 & 1.1 & 158.8 & 62.1 & 18.2 & -0.8 \\ \hline
2016 Jan 11 & PS1 & $z$ & 1 & 30 & 17.751 & 17.291 & 2.8 & 116.5 & 304.9 & 14.7 & +2.7 \\ \hline
2016 Feb 12 & PS1 & $w$ & 1 & 45 & 17.632 & 16.811 & 1.8 & 145.3 & 328.7 & 16.2 & +1.4 \\ \hline
2016 Mar 28 & PS1 & $r$ & 4 & 45 & 17.464 & 16.587 & 1.6 & 150.7 & 67.2 & 19.1 & -1.2 \\ \hline
2016 Apr 3 & PS1 & $w$ & 4 & 45 & 17.442 & 16.601 & 1.8 & 146.4 & 75.9 & 19.4 & -1.5 \\ \hline
2016 May 21 & PS1 & $i$ & 3 & 45 & 17.263 & 16.984 & 3.3 & 104.4 & 107.5 & 21.3 & -3.2 \\ \hline
2017 Apr 16 & SkyMapper & $g$ & 1 & 100 & 16.048 & 15.344 & 2.6 & 133.2 & 84.0 & 21.0 & -2.3 \\
&  & $r$ & 1 & 100 & & & & & & & \\
&  & $i$ & 1 & 100 & & & & & & & \\ \hline
2017 Apr 22 & SkyMapper & $r$ & 1 & 100 & 16.026 & 15.381 & 2.8 & 128.5 & 89.1 & 21.3 & -2.6 \\ \hline
2017 Apr 25 & PS1 & $w$ & 4 & 45 & 16.016 & 15.402 & 2.9 & 126.2 & 91.2 & 21.4 & -2.7 \\ \hline
2017 Apr 28 & PS1 & $w$ & 2 & 45 & 16.005 & 15.425 & 3.0 & 123.7 & 93.4 & 21.6 & -2.9 \\ \hline
2018 Mar 8 & PS1 & $w$ & 4 & 45 & 14.883 & 14.064 & 2.2 & 144.4 & 20.2 & 19.6 & +0.0 \\ \hline
2018 Apr 27 & ZTF & $g$ & 1 & 30 & 14.709 & 14.150 & 3.3 & 122.1 & 89.2 & 23.1 & -3.1 \\ \hline
2018 May 11 & ZTF & $g$ & 1 & 30 & 14.660 & 14.261 & 3.7 & 111.4 & 99.6 & 23.6 & -3.6 \\ \hline
2018 May 22 & ZTF & $r$ & 1 & 30 & 14.622 & 14.365 & 3.9 & 102.8 & 106.5 & 23.9 & -3.8 \\ \hline
2019 Apr 5 & PS1 & $w$ & 4 & 45 & 13.554 & 12.887 & 3.2 & 130.2 & 62.7 & 24.2 & -2.0 \\ \hline
2019 Jun 8 & DECam & $g$ & 1 & 90 & 13.347 & 13.320 & 4.4 & 89.3 & 118.7 & 26.6 & -4.3 \\ \hline
\enddata
\tablenotetext{a}{Heliocentric distance.}
\tablenotetext{b}{Observer-centric distance.}
\tablenotetext{c}{Phase angle.}
\tablenotetext{d}{Solar elongation.}
\tablenotetext{e}{Position angle of antisolar direction projected in the observer's plane of the sky.}
\tablenotetext{f}{Position angle of negative heliocentric velocity projected in the observer's plane of the sky.}
\tablenotetext{g}{Orbital plane angle. Negative values indicate the observer is below the orbital plane of the comet.}
\end{deluxetable*}

\begin{figure*}
\epsscale{1.1}
\begin{center}
\plotone{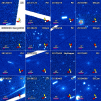}
\caption{
Collage of selected serendipitous archival observations of C/2019 E3 displayed in logarithmic scale at preperihelion heliocentric distances from $\sim\!23$ au to 13 au. The comet is placed at the centre of each panel. White stripes in some of the panels are CCD chip/cell gaps. Position angles of the antisolar direction and the negative heliocentric velocity projected in the sky plane are marked by the red and yellow arrows, respectively. Also shown are two scale bars, one with apparent lengths labelled, and the other one in magenta representing a linear length of $10^5$ km projected at the observer-centric distance of the comet. J2000 equatorial north points upwards and east to the left.
\label{fig:obs}
} 
\end{center} 
\end{figure*}

We primarily used the Solar System Object Image Search (SSOIS) tool \citep{2012PASP..124..579G} at the Canadian Astronomy Data Centre (CADC) to search for prediscovery serendipitous observations of C/2019 E3. In general, we were able to detect the comet in archival images from five different telescopes, Blanco 4 m, Pan-STARRS 1 (PS1), SkyMapper, VLT Survey Telescope (VST), and the Zwicky Transient Facility (ZTF), all the way back to 2012. Selected archival images of the comet are shown in Figure \ref{fig:obs}. In the following, we detail these observations separately based on the used telescopes.

\subsection{V{\'i}ctor M. Blanco 4 m Telescope}

Serendipitous archival observations of the comet in $g$, $i$, and $z$ filters from the Dark Energy Camera \citep[(DECam)][]{2015AJ....150..150F} on the V{\'i}ctor M. Blanco 4 m telescope at Cerro Tololo Inter-American Observatory, Chile, were identified. The camera houses 62 2k $\times$ 4k CCD chips at the prime focus, covering a hexagon-shaped 2\fdg2~diameter field-of-view (FOV) with an image scale of 0\farcs26 pixel$^{-1}$. The comet was located almost exactly at the ephemeris positions returned by JPL Horizons in the prediscovery archival DECam data on four different nights, plus one in 2019 somewhat over a year after the comet had been discovered (see Table \ref{tab:vgeo}). We estimated seeing by measuring the FWHM of field stars to be 1\farcs0-1\farcs3.

\subsection{Pan-STARRS}

Archival data from the two 1.8 m Pan-STARRS survey telescopes \citep{2016arXiv161205560C} on the summit of Haleakala, Maui, USA, including {\it w}-band images not publicly available, were searched based on ephemeris comparisons with the metadata of each image.  These telescopes have large 3\fdg1 FOVs with a grid of 60 CCD chips each further divided into $8 \times 8$ arrays of $590 \times 598$ pixel CCD cells having a 0\farcs25 pixel scale. Only PS1 observed the comet, in {\it w}-, {\it r}-, {\it i}-, and {\it z}-band images from 2012 to 2019, each of which has been astrometrically and photometrically calibrated \citep{2020ApJS..251....4W}. The FWHM of background stars varied between 0\farcs9 and 1\farcs6.

\subsection{SkyMapper}

Located at Siding Spring Observatory, Australia, SkyMapper is a 1.3 m survey telescope attached by a camera mosaicked by 32 CCD chips, each having $2048 \times 4096$ pixels at an image scale of 0\farcs50 pixel$^{-1}$, at the focal plane \citep{2007PASA...24....1K}. The serendipitous two-night multiband observations of C/2019 E3 from 2017 April returned by SSOIS were conveniently cropped to a dimension of $\sim\!10\arcmin \times 10\arcmin$ and centred on the target, which expediently assisted our identification of the object in prediscovery archival data. Neighbouring background stars have FWHM values between 2\farcs0 and 2\farcs3.

\subsection{VLT Survey Telescope}

We located comet C/2019 E3 in VST data from two nights in 2013 April obtained in $g$ and $r$ filters at the 2.6 m VST at Cerro Paranal, Chile, with OmegaCAM, which consists of 32 2k $\times$ 4k e2v CCD chips rendering an overall angular area of $\sim\!1 \times 1$ deg$^2$ at an angular resolution of 0\farcs21 pixel$^{-1}$ \citep{2002Msngr.110...15K}. As the observations returned by SSOIS were only raw images, we retrieved the corresponding bias and flat frames from the ESO Science Archive Facility and performed standard calibration. We measured the seeing during the observations, which varied between 1\farcs0 and 1\farcs2.

\subsection{Zwicky Transient Facility}

Together with SSOIS at CADC and the Moving Object Search Tool \citep{2019PASP..131a8003M} at IPAC\footnote{\url{https://irsa.ipac.caltech.edu/applications/MOST/}}, we collected fortuitous ZTF observations from three different nights in 2018 April to May in which the comet was found by visual inspection. These images at a pixel scale of 1\farcs01 were obtained by the wide-field ZTF camera, which is comprised of 16 e2v 6k $\times$ 6k CCDs covering a $\sim\!47$ deg$^{2}$ field of view on the 1.2 m Samuel Oschin Schmidt at the Palomar Observatory, USA, during its all-sky survey \citep{2019PASP..131a8002B,2019PASP..131g8001G,2019PASP..131a8003M}. Field stars in the images were measured to have seeing FWHM varying within a range of 2\farcs2-2\farcs9.

We tabulate detailed information of the serendipitous archival observations along with the viewing geometry of the comet in Table \ref{tab:vgeo}.

\section{Results}
\label{sec_rslt}

\subsection{Morphology}
\label{ss_morph}

Visually inspecting the observed morphology of C/2019 E3 in the serendipitous archival data, we found that the comet displayed no prominent tail but maintained a symmetric appearance in general (see Figure \ref{fig:obs}), regardless of the orbital plane angle at which it was observed (see Table \ref{tab:vgeo}). In particular, this symmetric morphology largely remained even for the PS1 data obtained from 2018 March 8 when the Earth was practically within the orbital plane of the comet, although only one image from the imaging sequence managed to escape from an artefact caused by bleeding of a nearby bright star. Only in the post-discovery DECam data taken from 2019 June 8 could we notice hints of elongation in the appearance of the comet (see Figure \ref{fig:obs}). The insensitivity of the comet's morphology to the changing orbital plane angle implies that its observed dust environment was dominated by relatively large-sized particles, as they are less susceptible to solar radiation pressure than their smaller counterparts. Otherwise, we would expect the comet to be extended considerably in its orbital plane in comparison to the out-of-plane direction. 

\begin{figure*}
\begin{center}
\gridline{\fig{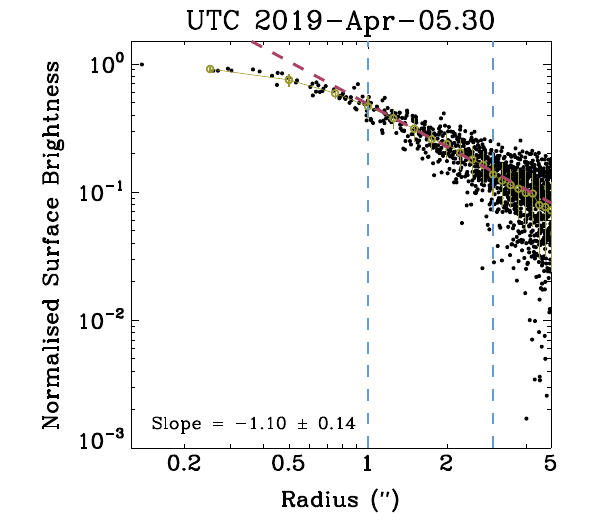}{0.5\textwidth}{(a)}
\fig{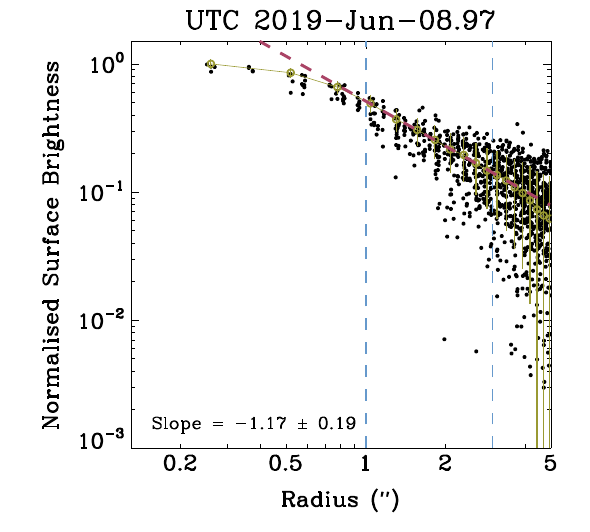}{0.5\textwidth}{(b)}}
\caption{
Normalised radial surface brightness profiles of C/2019 E3 from (a) 2019 April 5 in the PS1 data and (b) 2019 June 8 in the DECam data. Azimuthally mean values are plotted in olive, with the best-fit power-law radial profiles plotted as pink dashed lines. In each panel, the inner and outer radii of the fitted annular region are marked by two vertical blue lines, and the value of the best-fit logarithmic surface brightness slope is given in the lower left corner.
\label{fig:rprof}
} 
\end{center} 
\end{figure*}

We examined the radial brightness profile of the comet in 2019 when it had the best signal-to-noise ratios in the available archival data. A power-law surface brightness model was fitted to the observations in an annular region between 1\arcsec~and 3\arcsec~in radius from the comet after the sky background was removed, whereby we obtained the best-fit logarithmic surface brightness slope of the comet. The values were found to be statistically consistent with the one in steady state (see Figure \ref{fig:rprof}). We verified that adjusting the fitted region did not affect the results beyond the noise level. Therefore, we are confident to conclude that the mass loss of the comet was most likely in steady state, suggestive of activity driven by sublimation. Given that temperature at the observed range of heliocentric distances would be too low for water ice to sublimate, we prefer sublimation of supervolatiles such as carbon monoxide (CO) and carbon dioxide (CO$_{2}$) as the activity mechanism at play.

The observed morphology of C/2019 E3 has a close resemblance to that of C/2017 K2 (PANSTARRS), another ultra-distant comet whose dust coma was reported to be primarily comprised of submillimetre-scaled grains ejected in a protracted manner at speeds of $\la\!3$ m s$^{-1}$ at similar heliocentric distances \citep{2018AJ....155...25H,2019AJ....157...65J}. Therefore, we conjectured that the physical properties of the dust environment of C/2019 E3 highly resembled that of C/2017 K2. In Section \ref{ss_dust}, we will detail the application of our Monte Carlo dust model to investigate the physical properties of the dust environment of C/2019 E3.

\subsection{Photometry}
\label{ss_phot}

\begin{figure}
\epsscale{1.2}
\begin{center}
\plotone{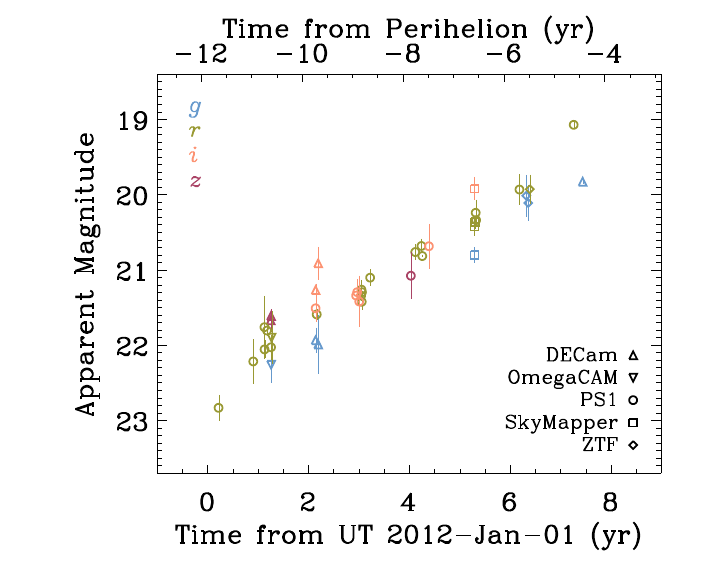}
\caption{
Apparent magnitude of comet C/2019 E3 in multiple bands in the SDSS system measured with a $3\times10^{4}$ km radius aperture versus time. Data points are colour coded according to the calibrated photometric bands and plotted in different symbols representing different facilities as shown in the legends. In general, the apparent brightness of the comet increased steadily as it approached perihelion.
\label{fig:lc_app}
} 
\end{center} 
\end{figure}

\begin{figure}
\epsscale{1.2}
\begin{center}
\plotone{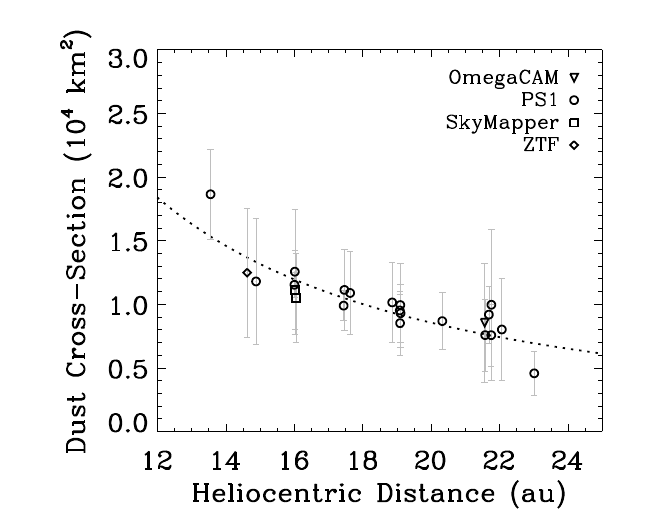}
\caption{
Effective scattering cross-section of dust as a function of heliocentric distance, calculated from $r$-band data points measured with an aperture of $3 \times 10^{4}$ km in radius. Data points in different symbols correspond to measurements from different facilities. The dotted curve is the best-fit power law with heliocentric distance for the effective scattering cross-section of dust. The overall trend is that the effective scattering cross-section of dust increased monotonically as the comet approached the Sun. 
\label{fig:xs_r}
} 
\end{center} 
\end{figure}

The serendipitous archival images containing C/2019 E3 were photometrically calibrated using ATLAS Refcat2 \citep{{2018ApJ...867..105T}}. We transformed measurements in the PS1 photometric system to the SDSS system following \citet{2012ApJ...750...99T}. Photometry of C/2019 E3 in archival data from the five telescopes was carried out using circular apertures having fixed linear radii from $2.5 \times 10^{4}$ to $4 \times 10^{4}$ km at a step size of 5000 km projected at the distance of the comet. The benefit of utilising such apertures is that, in spite of the observer-centric distance of the comet varying as a function of time, the sampled region around the comet remained fixed, thereby avoiding the introduction of unnecessary influences due to the changing viewing geometry as much as possible. Figure \ref{fig:lc_app} shows the apparent magnitude of the comet enclosed by an aperture of $3 \times 10^{4}$ km in radius against time in the corresponding SDSS-system bandpasses, in which the magnitude uncertainties were standard deviations of multiple measurements from the same telescopes and filters in the same nights, or, in case of only single measurements available, propagated from Poisson statistics and errors in zero-points of images. During the observed period, the comet basically steadily brightened on its way to perihelion, exhibiting no compelling evidence of outbursts in brightness.

The observed steady brightening of C/2019 E3 may imply its gradually increasing total effective scattering cross-section of dust. We computed this quantity using the $r$-band data points of our measurements through the following equation,
\begin{equation}
{\it \Xi}_{\rm e} = \frac{\pi}{p_{r} \phi \left(\alpha\right)}\left(\frac{r_{\rm H} {\Delta}}{r_{\oplus}} \right)^{2} 10^{0.4 \left(m_{\odot,r} - m_r \right)}
\label{eq_XS},
\end{equation}
\noindent where $m_r$ is the apparent $r$-band magnitude of the comet, $m_{\odot,r} = -26.93$ is the apparent $r$-band magnitude of the Sun at mean heliocentric distance of Earth $r_{\oplus} = 1$ au \citep{2018ApJS..236...47W}, $r_{\rm H}$ and $\Delta$ are respectively heliocentric and observer-centric distances of the comet, and $\phi$ and $p_{r}$ are, respectively, the dimensionless phase dependency as a function of phase angle $\alpha$ and the $r$-band geometric albedo of dominant dust grains comprising the coma of the comet. As there is no observation that can possibly provide constraints on the latter two quantities, we had to assume a nominal value of $p_r = 0.05$ for the geometric albedo \citep[e.g.,][]{2018SSRv..214...64L} and adopt a linear phase function with typical phase coefficient $\beta_{\alpha} = 0.03 \pm 0.01$ mag degree$^{-1}$ \citep{1987A&A...187..585M} for optically dominant dust grains in the coma. The result is plotted against heliocentric distance in Figure \ref{fig:xs_r}, which shows that the effective scattering cross-section of dust generally increased as the comet approached the Sun, as expected. We used the power-law form of ${\it \Xi}_{\rm e} \sim r_{\rm H}^{\gamma}$, where $\gamma$ is a constant power-law index, to least-squared fit the dataset. The best-fit power-law index for the heliocentric distance dependency is $\gamma = -1.5 \pm 0.4$, where the stated uncertainty is the $1\sigma$ formal error propagated from the counterparts in the measurements. We note that the obtained value is not statistically different from the one for C/2017 K2, another ultra-distant comet, at a similar range of heliocentric distances \citep[$\gamma = -1.14 \pm 0.05$;][]{2021AJ....161..188J}. On the other hand, the behaviour of C/2019 E3 was dissimilar to that of the other two known ultra-distant comets C/2010 U3 and C/2014 UN$_{271}$, both of which exhibited obvious outbursts at similar preperihelion heliocentric distances \citep{2019AJ....157..162H,2022ApJ...933L..44K}, rendering the power-law model inapplicable.

\begin{deluxetable*}{ccCCCCC}
\tablecaption{Colour Measurements of Comet C/2019 E3 (ATLAS)
\label{tab:clr}}
\tablewidth{0pt}
\tablehead{
\colhead{Date (UTC)} & Facility & \multicolumn{5}{c}{Colour Measurement} \\
\cmidrule(lr){3-7}
&& \colhead{Colour} & \multicolumn{4}{c}{Photometric Aperture Radius ($10^4$ km)} \\
\cmidrule(lr){4-7}
&& \colhead{Index} & 2.5 & 3.0 & 3.5 & 4.0
}
\startdata
2013 Apr 3 & DECam \& PS1 & r-z & +0.41 \pm 0.17 & +0.44 \pm 0.09 & +0.52 \pm 0.08 & +0.55 \pm 0.13 \\ 
2014 Feb 20 & DECam & g-i & +0.67 \pm 0.19 & +0.68 \pm 0.20 & +0.71 \pm 0.20 & +0.67 \pm 0.29 \\ 
2014 Mar 11 & DECam & g-i & +1.08 \pm 0.44 & +1.05 \pm 0.43 & +0.98 \pm 0.48 & +1.03 \pm 0.25 \\ 
2017 Apr 16 & SkyMapper & g-r & +0.38 \pm 0.17 & +0.42 \pm 0.16 & +0.41 \pm 0.15 & +0.32 \pm 0.16 \\
&& r-i & +0.51 \pm 0.20 & +0.39 \pm 0.20 & +0.30 \pm 0.20 & +0.42 \pm 0.21 \\
&& g-i & +0.89 \pm 0.19 & +0.82 \pm 0.19 & +0.71 \pm 0.20 & +0.73 \pm 0.19 \\ \hline
\multicolumn{2}{l}{Weighted Mean} & g-r & \multicolumn{4}{c}{$+0.38 \pm 0.05$} \\
&& g-i & \multicolumn{4}{c}{$+0.78 \pm 0.13$} \\
&& r-i & \multicolumn{4}{c}{$+0.41 \pm 0.08$} \\
&& r-z & \multicolumn{4}{c}{$+0.49 \pm 0.06$}
\enddata
\tablecomments{The reported uncertainties are weighted standard deviations of repeated measurements. Weighted mean colours of the comet were computed from measurements using different photometric apertures. For comparison, here we list also the median colours of comets reported by \citet{2012Icar..218..571S}, $g - r = +0.57 \pm 0.05$, $r - i = +0.22 \pm 0.07$, and $i - z = +0.09 ± 0.07$.}
\end{deluxetable*}

The multiband observations of the comet allowed for measurements of its colour at several epochs (see Table \ref{tab:clr}). Due to the large uncertainty in the photometric measurements, we found no compelling evidence of radial gradients in the colour of the coma. Unfortunately, only the $g - i$ colour of the comet was available from multiple epochs, which are highly unevenly spaced in time -- the first two were obtained from two nights merely a month apart from each other, and the third one was measured over three years later after the first pair. From these measurements, we saw no evidence of temporal variation in the $g - i$ colour of the comet statistically beyond the noise level. Therefore, we computed the weighted mean colour indices of the comet from the repeated photometric measurements regardless of the used photometric apertures. We obtained $g - i = +0.78 \pm 0.13$, $g - r = +0.38 \pm 0.05$, $r - i = +0.41 \pm 0.08$, and $r - z = +0.49 \pm 0.06$ (See also Table \ref{tab:clr}). In comparison to the solar colours, $g - r = +0.46 \pm 0.03$, $r - i = +0.12 \pm 0.03$, $r - z = +0.15 \pm 0.03$, which we derived from \citet{2018ApJS..236...47W}, while the colour of the comet in $g - r$ appeared to be similar to that of the Sun given the uncertainty, at longer wavelengths the comet was likely redder than the Sun. We also compared the colour of C/2019 E3 to those of other long-period comets as well as short-period ones, reported by \citet{2012Icar..218..571S}, finding that while the comet appeared to have a somewhat bluer colour in $g - r$ than typical comets, the colour of the comet in other wavelength intervals are fully comparable.

\subsection{Orbit Determination}
\label{ss_orb}

\begin{deluxetable}{lcc}
\tablecaption{Best-fitted Orbital Solution for Comet C/2019 E3 (ATLAS)
\label{tab:orb}}
\tablewidth{0pt}
\tablehead{
\multicolumn{2}{c}{Quantity}  & 
Value
}
\startdata
Eccentricity & $e$
       & 0.9986942(40) \\ 
Perihelion distance (au) & $q$
       & 10.312981(15) \\ 
Semimajor axis ($10^{3}$ au) & $a$
       & 7.897(24) \\
Inclination (\degr) & $i$
       & 84.2989171(87) \\ 
Argument of perihelion (\degr) & $\omega$
                 & 280.69964(14) \\ 
Longitude of ascending node (\degr) & ${\Omega}$
                 & 347.2325934(49) \\ 
Time of perihelion (TT)\tablenotemark{$\dagger$} & $t_\mathrm{p}$
                  & 2023 Nov 15.2794(30) \\
\hline
\multicolumn{2}{l}{Observed arc}
& 2012 Mar 19-2023 Dec 11\\
\multicolumn{2}{l}{Number of observations used (discarded)}
& 675 (7) \\
\multicolumn{2}{l}{Residual rms (\arcsec)}
& 0.435 \\
\multicolumn{2}{l}{Normalised residual rms}
& 0.570
\enddata
\tablenotetext{\dagger}{The uncertainty is in days.}
\tablecomments{The osculating orbit is referenced to the heliocentric J2000 ecliptic at epoch TT 2023 Dec 11.0 = JD 2460289.5. Here, numbers in parentheses of the orbital elements are $1\sigma$ formal errors of the corresponding parameters.}
\end{deluxetable}

We exploited field stars to solve plate constants of the serendipitous archival data referenced to the Gaia Data Release 2 and 3 catalogues \citep{2018A&A...616A...1G, 2023A&A...674A...1G}, during which process the field stars were simply treated as bidimensional symmetric Gaussians to be fitted. Despite that none of the archival data tracked the apparent nonsidereal motion of the comet, the motion was slow enough that the optocentres of the comet remained circularly symmetric enough. Thus, we also simply treated the comet as a bidimensional symmetric Gaussian, whereby the best-fitted pixel coordinates of the centroid of the comet were obtained. We then transformed the pixel coordinates of the comet to the J2000 equatorial coordinate system in terms of R.A. and Decl. Meanwhile, the corresponding uncertainties were obtained by propagating the counterparts in centroiding and astrometric calibration, based upon which our astrometry was properly weighted. In addition, we included more recent astrometric measurements of the comet returned by the Minor Planet Center Database Search\footnote{\url{https://minorplanetcenter.net/db_search}}. As the dataset from the Minor Planet Center contained no information on the measurement uncertainties and consisted of astrometric reduction to a mix of various star catalogues, we debiased and assigned a weighting scheme for the data following descriptions detailed in \citet{2020Icar..33913596E} and \citet{2017Icar..296..139V}, respectively. We then fed the astrometric observations with the adopted weighting scheme into the orbit determination package {\tt Find\_Orb}\footnote{The orbit determination package is developed by B. Gray, publicly available at \url{https://github.com/Bill-Gray/find_orb}}, which incorporated gravitational perturbations from the eight major planets, Pluto, the Moon, and the 16 most massive asteroids in the main belt as well as relativistic effects. Planetary and lunar ephemeris DE440 \citep{2021AJ....161..105P} was exploited to speed up the N-body integration process of the package. Our measurements with observed-minus-calculated ($O-C$) astrometric residuals greater than the $3\sigma$ level were slightly downweighted accordingly, whereas six of the measurements from the Minor Planet Center were simply discarded, because the latter all have $O-C$ residuals greater than a few arcseconds, at least an order of magnitude worse than the former. We tabulate the best-fitted osculating heliocentric orbital elements in Table \ref{tab:orb}. It is worth pointing out that the current perihelion distance of the comet is the third largest for known comets after C/2003 A2 (Gleason) and C/2014 UN$_{271}$ (with $q = 11.4$ au and 10.9 au, respectively, according to solutions by JPL Horizons).

\section{Discussion}
\label{sec_disc}

\subsection{Dust Properties}
\label{ss_dust}

\begin{figure*}
\epsscale{1.}
\begin{center}
\plotone{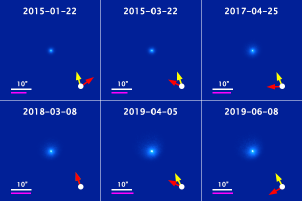}
\caption{
Monte Carlo dust models of comet C/2019 E3 at six of the selected epochs. All of the images are scaled logarithmically. In each panel, the red and yellow arrows mark the position angles of the antisolar direction and the negative heliocentric velocity projected in the plane of the sky, respectively, and two scale bars of 10\arcsec~in apparent length (labelled) and $10^5$ km projected at the observer-centric distance of the comet are shown in white and magenta, respectively. J2000 equatorial north is up and east is left.
\label{fig:dust_mdl}
} 
\end{center} 
\end{figure*}

We applied our Monte Carlo dust dynamical code to simulate the dust morphology of C/2019 E3 from a list of epochs observed in good quality at various orbital plane angles, as a way to probe the physical properties of its dust environment. The model assumed that dust grains of spherical shape were ejected from the nucleus due to sublimative activity, forming a cone-shaped jet symmetric about the Sun-comet axis towards the Sun, at a range of nonzero terminal ejection speeds following an empirical relation of \citep[e.g.,][]{2008Icar..193...96I}
\begin{equation}
V_{\rm ej} = V_{\rm ej,0} \sqrt{\left(\frac{r_{\oplus}}{r_{\rm H}}\right) \left(\frac{\mathfrak{a}_{\rm d,0}}{\mathfrak{a}_{\rm d}} \right)}
\label{eq_Vej}.
\end{equation}
\noindent Here, $\mathfrak{a}_{\rm d}$ is the grain radius, $\mathfrak{a}_{\rm d,0} = 5$ mm is the referenced grain radius, and $V_{\rm ej,0}$ is the terminal ejection speed of reference-sized dust at $r_{\oplus}$ from the Sun. For simplicity, the acceleration process of dust grains to terminal speeds was ignored, such that they instantaneously reached terminal speeds after ejection. The dimension of the nucleus was ignored in the model as well. Afterward, each of the dust grains was considered to be dominantly subject to the solar radiation pressure force and the solar gravitational force, whose ratio, denoted by $\beta_{\rm rp}$, is inversely proportional to $\rho_{\rm d} \mathfrak{a}_{\rm d}$ ($\rho_{\rm d}$ is the bulk density of dust, assumed to be 1 g cm$^{-3}$). The gravity of the nucleus was neglected. We used a power-law size distribution for the number of dust grains, i.e., ${\rm d} \mathcal{N} \propto \mathfrak{a}_{\rm d}^{-\gamma} {\rm d}\mathfrak{a}_{\rm d} $ (${\rm d}\mathcal{N} $ is the number of dust grains having radii from $\mathfrak{a}_{\rm d}$ to $\mathfrak{a}_{\rm d} + {\rm d}\mathfrak{a}_{\rm d}$, and $\gamma$ is the power-law index of the size distribution), in a dust-size range of $\mathfrak{a}_{\rm d, \min} \le \mathfrak{a}_{\rm d} \le \mathfrak{a}_{\rm d, \max}$. The production rate was assumed to be inversely proportional to the square of the heliocentric distance. Motions of the simulated ejected particles alongside the nucleus itself were integrated from the time of ejection to the corresponding observed epochs in our implemented version of {\tt mercury6} \citep{1999MNRAS.304..793C}. Our code then transformed the Cartesian heliocentric states of the nucleus and particles to topocentric ones, with light-travel time corrected. After adopting the image scales of the actual observations selected for modelling, we thereby attained model images of the comet to be compared against actual observations. Earlier versions of our dust dynamical code have been previously applied for various active small solar system bodies including another ultra-distant comet C/2017 K2 \citep{2018AJ....155...25H,2019AJ....157...65J}.

Unfortunately, owing to the faintness of the comet and the quality of the archival data, we could not simply treat all the relevant physical parameters of the dust environment as free parameters to be solved by straightforwardly fitting the morphology of the comet. Rather, we started with physical parameters of dust grains similar to those of comet C/2017 K2 in \citet{2018AJ....155...25H} and \citet{2019AJ....157...65J} and tweaked them manually when necessary. Given our earliest archival observation of the comet in 2012, we set the onset time of activity to be early 2011. We have verified that adopting even earlier epochs or slightly later epochs had no appreciable effect in the resulting modelled morphology. The maximum grain size and the power-law index of the dust-size distribution were found to have no significant effect on the modelled dust morphology either, as long as the latter parameter satisfies $\gamma > 3$, which holds for the great majority of comets \citep{2004come.book..565F}, and so we simply adopted $\mathfrak{a}_{\rm d, \max} = 1$ cm and a nominal value of $\gamma = 3.5$ \citep[e.g.,][]{2023arXiv230912759A}. On the contrary, the modelled morphology is mostly sensitive to the input minimum dust size and ejection speeds, as expected. By employing step sizes of 1 mm and 0.1 m s$^{-1}$ for $\mathfrak{a}_{\rm d, \min}$ and $V_{\rm ej, 0}$, respectively, we ended up obtaining models with particles having $\mathfrak{a}_{\rm d, \min} = 2$ mm and $V_{\rm ej, 0} = 1.2$ m s$^{-1}$ released from the sunlit hemisphere of the nucleus starting since 2011 vividly reproducing the observed morphology of the comet (see Figure \ref{fig:dust_mdl}, to be compared with observations in Figure \ref{fig:obs}). Scaling with the obtained reference ejection speed using Equation (\ref{eq_Vej}), our model suggests that the observed dust morphology of the comet was likely formed by protracted ejections of dust of mm-scale and greater at speeds $\la\!0.4$ m s$^{-1}$ in the observed range of heliocentric distances. As a comparison, \citet{2018AJ....155...25H} and \citet{2019AJ....157...65J} reported that the optically dominant dust in the coma of C/2017 K2 was at least submillimetre scaled and had ejection speeds $\la\!4$ m s$^{-1}$ at similar heliocentric distances. As such, the results from our Monte Carlo dust modelling for C/2019 E3 are in line with the scenario where the observed activity of the comet is driven by sublimation of supervolatiles, the same activity mechanism as for C/2017 K2.

\subsection{Activity}
\label{ss_act}

\begin{figure}
\epsscale{1.2}
\begin{center}
\plotone{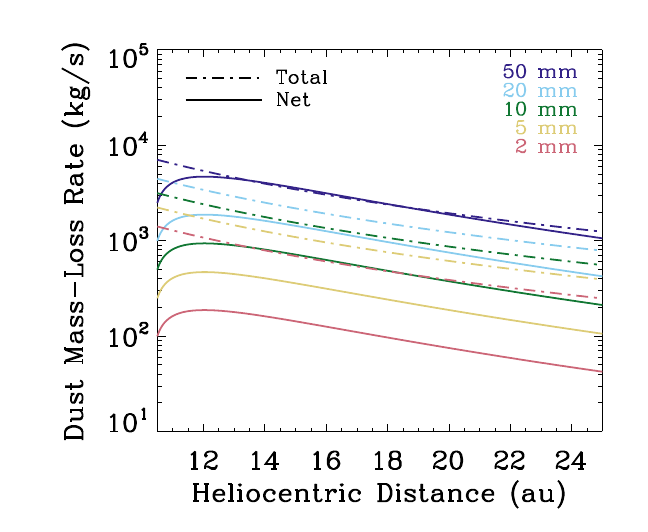}
\caption{
Total and net mass-loss rates of comet C/2019 E3 versus heliocentric distance (dashed-dotted and solid curves, respectively) assuming different effective radii of dust grains (plotted in different colours). The best-fit effective scattering cross section of dust in the power-law form versus heliocentric distance (see Section \ref{ss_phot}) was used.
\label{fig:mloss}
} 
\end{center} 
\end{figure}

\begin{figure*}
\begin{center}
\gridline{\fig{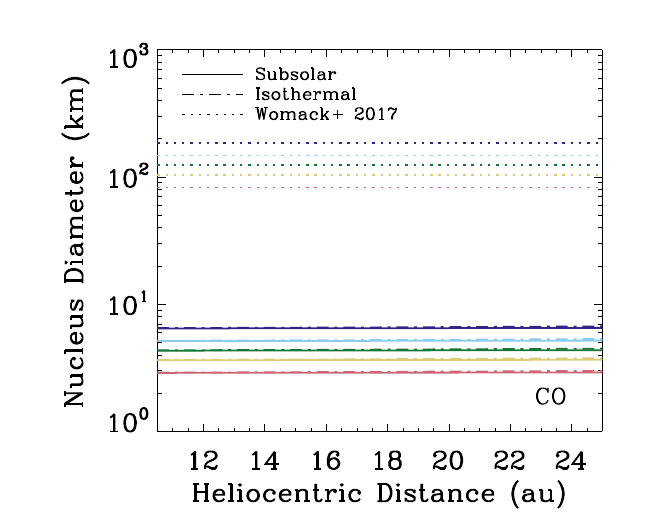}{0.5\textwidth}{(a)}
\fig{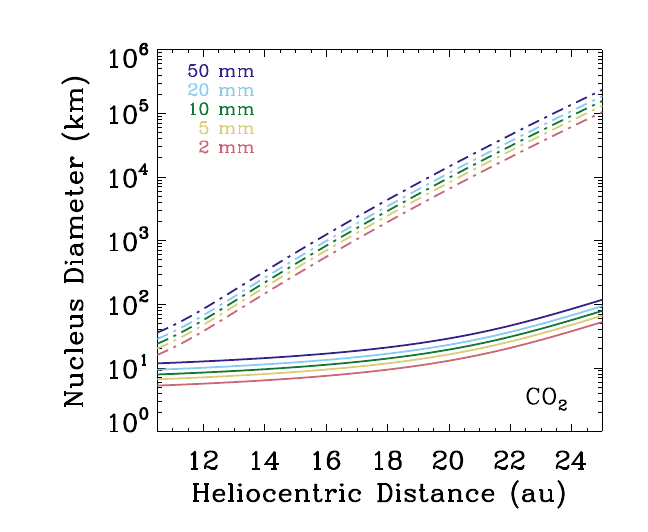}{0.5\textwidth}{(b)}}
\caption{Lower limits to the nucleus diameter of C/2019 E3 as functions of heliocentric distance assuming the observed activity is driven by sublimation of CO (a) and CO$_{2}$ (b). Results in subsolar and isothermal scenarios are plotted as solid and dashed-dotted curves, respectively, in different colours representing different effective sizes of dust grains used for the computation. We conservatively assumed $\mathcal{X} = 5$ for the ratio of the dust-to-gas mass production rates in the calculation. In addition, we adopted an empirical model by \citet{2017PASP..129c1001W} that is consistent with the CO measurements of comet C/1995 O1 (Hale-Bopp) and Centaur 29P/Schwassmann-Wachmann in the left panel, extrapolating it to greater heliocentric distances.
\label{fig:nuc_rad}
} 
\end{center} 
\end{figure*}

The total mass-loss rate of C/2019 E3 driven by steady-state ejection of dust is given by the total mass within the region enclosed by some photometric aperture divided by the aperture crossing time of dust, i.e.,
\begin{equation}
\dot{\mathcal{M}}_{\rm d} \approx \frac{4 \rho_{\rm d} \overline{\mathfrak{a}}_{\rm d} \overline{V}_{\rm ej} {\it \Xi}_{\rm e}}{3 \ell}
\label{eq_mloss}.
\end{equation}
\noindent Here, $\overline{\mathfrak{a}}_{\rm d}$, and $\overline{V}_{\rm ej}$ are respectively the mean radius and mean ejection speed of dust particles in the coma, and $\ell$ is the aperture radius projected at the distance of the comet. Unfortunately, because we could not robustly determine the maximum size and size-distribution distribution of ejected dust (Section \ref{ss_dust}), we instead varied the mean dust radius from the minimum size and scaled the corresponding ejection speed using Equation (\ref{eq_Vej}). The results with the best-fit power-law form of the effective scattering cross-section of dust as a function of heliocentric distance are plotted in Figure \ref{fig:mloss}. Also plotted are net mass-loss rates of dust having various mean dust radii, which we calculated from
\begin{align}
\nonumber
\frac{\Delta \mathcal{M}_{\rm d}}{\Delta t} 
\approx & \frac{4}{3} \rho_{\rm d} \overline{\mathfrak{a}}_{\rm d} \frac{\Delta {\it \Xi}_{\rm e}}{\Delta t} \\
\begin{split}
\label{eq_nmloss_1}
= & -\frac{4}{3} \rho_{\rm d} \overline{\mathfrak{a}}_{\rm d} \left(\frac{\Delta {\it \Xi}_{\rm e}}{\Delta r_{\rm H}}\right) \frac{1}{r_{\rm H}} \\
& \cdot \sqrt{\mu_{\odot} \left(\frac{r_{\rm H} - q}{q}\right) \left[\left(1 + e \right) q - \left(1 - e \right) r_{\rm H} \right]} 
\end{split}
\\
\approx & -\frac{4}{3}\rho_{\rm d} \overline{\mathfrak{a}}_{\rm d} \left(\frac{\Delta {\it \Xi}_{\rm e}}{\Delta r_{\rm H}}\right) \frac{\sqrt{2 \mu_{\odot} \left(r_{\rm H} - q\right)}}{r_{\rm H}}
\label{eq_nmloss_2}.
\end{align}
\noindent Here, $\mu_{\odot} \triangleq G M_{\odot}$, in which $G$ is the gravitational constant and $M_{\odot}$ is the mass of the Sun, is the mass parameter of the Sun, and we applied the chain rule to arrive at Equation (\ref{eq_nmloss_1}), which can be further approximated to Equation (\ref{eq_nmloss_2}) with a near-parabolic eccentricity of $e \approx 1$. Given the unknowns and uncertainties in the pertinent physical parameters of dust grains, the results from these calculations shown in Figure \ref{fig:mloss} are likely no better than order-of-magnitude estimates. During the observed timespan, the comet experienced a total dust mass-loss rate of $\ga\!10^{2}$ kg s$^{-1}$ and net mass-loss rate of $\ga \! 10$ kg s$^{-1}$ even at heliocentric distance $r_{\rm H} \ga 20$ au. We note that such a level of activity was comparable to that of C/2017 K2 (PANSTARRS) at similar distances from the Sun \citep[total mass-loss rate $\ga\!200$ kg s$^{-1}$; ][]{2018AJ....155...25H,2019AJ....157...65J}.

In the following, we proceed to estimate the minimum size of the nucleus that would be required to sustain the activity of C/2019 E3. Given the great heliocentric distances of the comet, the protracted mass production is mostly consistent with activity driven by sublimation of supervolatiles such as CO and CO$_{2}$, as in the cases of the other three ultra-distant comets. Adopting a ratio of dust-to-gas mass production rates of $\mathcal{X}$, we can relate the minimum sublimating area to the total mass-loss rate of dust as
\begin{equation}
\mathcal{A}_{\rm s} = \frac{\dot{\mathcal{M}_{\rm d}}}{\mathcal{X} f_{\rm s}}
\label{eq_As}.
\end{equation}
\noindent Here, $f_{\rm s}$ is the mass flux of some sublimating substance, which can be numerically solved from the following energy equilibrium equation
\begin{equation}
\left(1 - A_{\rm B} \right) S_{\odot} \left(\frac{r_{\oplus}}{r_{\rm H}}\right)^{2} \cos \zeta = \epsilon \sigma T^{4} + L\left(T \right) f_{\rm s} \left(T\right)
\label{eq_E_conserved}.
\end{equation}
\noindent The left-hand side represents insolation at the nucleus, which is converted to energy spent in thermal reradiation and sublimation (the two terms on the right-hand side). In the above equation, $A_{\rm B}$ is the Bond albedo of the nucleus, $S_{\odot} = 1361$ W m$^{-2}$ is the solar constant, $\cos \zeta$ is the effective projection coefficient for the surface in a range of $\cos \zeta \in \left[1/4, 1 \right]$, with the upper and lower bounds corresponding to subsolar and isothermal scenarios, respectively, $\sigma = 5.67 \times 10^{-8}$ W m$^{-2}$ K$^{-4}$ is the Stefan-Boltzmann constant, $\epsilon$ and $T$ are the surface emissivity and equilibrium temperature, respectively, and $L$ is the latent heat of the sublimating substance varying with temperature. To solve Equation (\ref{eq_E_conserved}) for the mass fluxes of outgassing CO and CO$_{2}$, we assumed a conservative ratio of dust-to-gas mass production rates $\mathcal{X} = 5$ as in \citet{2019AJ....157...65J}, $A_{\rm B} = 0.01$, and $\epsilon = 0.9$, both of which are typical for cometary nuclei \citep{2017Icar..284..344K,2023arXiv230409309K}, and adopted the empirical sublimation models by \citet{2009P&SS...57.2053F}. In the subsolar scenario, the minimum diameter of the cometary nucleus is simply the diameter of the equal-area circle, whereas in the isothermal case, it is given by the diameter of a sphere having the same surface area as the minimum sublimating area yielded by Equation (\ref{eq_As}). We plot the results for CO and CO$_{2}$ respectively in the left and right panels of Figure \ref{fig:nuc_rad}, in which we immediately notice that the isothermal sublimation models of CO$_{2}$ should be best rejected, in that they all predict unreasonably enormous nucleus sizes for the comet. On the contrary, results from the subsolar model with CO$_{2}$ are not different appreciably from those from the subsolar and isothermal models with CO, which themselves are nearly indistinguishable from each other. Aside from these models, we also adopted an empirical model by \citet{2017PASP..129c1001W} that is consistent with the CO measurements of comet C/1995 O1 (Hale-Bopp) and Centaur 29P/Schwassmann-Wachmann at heliocentric distances from $\sim\!4$ to 10 au. Extrapolating the empirical model to greater heliocentric distances, we find that the estimated diameter for the nucleus of C/2019 E3 is $\sim\!80$ km, at least an order of magnitude greater than our results with the sublimation models by \citet{2009P&SS...57.2053F}. The discrepancy may imply that either the nucleus of the comet is even larger than those of C/1995 O1 and 29P, which were reported to have effective nucleus radii of $\sim\!37$ km and 30 km, respectively \citep{2012ApJ...761....8S,2015Icar..260...60S,2021PSJ.....2..126S}, or that the latter two objects would be less active than was C/2019 E3 at heliocentric distances $r_{\rm H} \ga 10$ au. On the whole, we are confident to conclude that the nucleus of C/2019 E3 is most likely greater than $\sim\!3$ km in diameter.


\subsection{Orbital Evolution}
\label{ss_orbev}

\begin{deluxetable*}{lccc}
\tablecaption{Original and Future Orbits of Comet C/2019 E3 (ATLAS)
\label{tab:orb_OF}}
\tablewidth{0pt}
\tablehead{
\multicolumn{2}{c}{Quantity}  & 
Original & Future
}
\startdata
Eccentricity & $e$
       & 0.9997085(40)
       & 0.9994176(40) \\ 
Periapsis distance (au) & $q$
       & 10.313058(15)
       & 10.308713(15) \\ 
Semimajor axis ($10^{4}$ au) & $a$
       & 3.539(48)
       & 1.770(12) \\
Reciprocal of semimajor axis ($10^{-5}$ au$^{-1}$) & $a^{-1}$
       & 2.826(39)
       & 5.649(39)\\
Inclination (\degr) & $i$
       & 84.2820939(87)
       & 84.2848057(87) \\ 
Argument of periapsis (\degr) & $\omega$
                 & 280.70932(14)
                 & 280.68641(14) \\ 
Longitude of ascending node (\degr) & ${\Omega}$
                 & 347.1826022(48)
                 & 347.1801992(48) \\ 
Time of periapsis (TDB)\tablenotemark{$\dagger$} & $t_\mathrm{p}$
                  & 2023 Nov 16.4211(30)
                  & 2023 Nov 16.7034(30) \\
\hline
\multicolumn{2}{l}{Epoch (TDB)\tablenotemark{$\dagger$}} & 
    1709 May $20.8 \pm 1.7$ & 
    2338 Sep $19.8 \pm 1.7$
\enddata
\tablenotetext{\dagger}{The uncertainties are in days.}
\tablecomments{Both of the original and future orbits are referred to the solar system barycentre in the J2000 ecliptic coordinate system at epochs when each of the clones is at pre- and post-perihelion heliocentric distances 250 au, respectively. The uncertainties of the orbital elements are standard deviations computed from distributions of the 5,001 clones.}
\end{deluxetable*}

It is of scientific interest to understand why C/2019 E3 could be so active even at great distances from the Sun and whether this behaviour implies that it is one of the most pristine objects in our solar system. As the thermophysical status of the comet is closely associated with its dynamical history, we investigated its orbital evolution and examined whether or not C/2019 E3 has previously entered the planetary region, i.e., being dynamically old or new, and if the observed activity could be attributed to retained heat from the previous perihelion return. We created 5,000 Monte Carlo clones of the nominal orbit based on the best-fit orbital elements and the covariance matrix obtained from the orbit determination detailed in Section \ref{ss_orb}, all of which, together with the nominal orbit, were integrated backward in time using our modified version of {\tt mercury6} until they reach preperihelion heliocentric distance $r_{\rm H} = 250$ au, beyond which planetary perturbations are negligible. The osculating orbit when at preperihelion heliocentric distance $r_{\rm H} = 250$ au is termed the ``original" orbit of the comet, following the definition by \citet{2001A&A...375..643D}. The N-body integration took into account perturbations from the eight major planets, Pluto, the Moon, and the 16 most massive asteroids in the main belt, the heliocentric states of which were taken from DE440. Also included in the force model were post-Newtonian corrections and the Galactic tide assuming a local stellar density of $\rho_{\ast} = 0.1 M_{\odot}$ pc$^{-3}$ in the solar neighbourhood and using the formalism detailed in Appendix \ref{app_eom}, which was derived based on \citet{2005CeMDA..93..229F}. We list the orbital elements and the corresponding uncertainties of the original orbit of C/2019 E3 computed from the 5,001 orbital clones in Table \ref{tab:orb_OF}, referred to the solar system barycentre in the J2000 ecliptic coordinate system. Then starting with the obtained original orbit, assuming gravitational perturbations from nearby passing stars are unimportant, we adopted an analytic approach to evaluate the change in periapsis distance of the comet between the previous and current returns in the solar system barycentric reference system.

The Hamiltonian for a comet orbiting around the barycentre of the solar system under the influence of the tidal potential from the Galactic disc is 
\begin{equation}
H = -\frac{\mu_{0}}{2a} + \underbrace{2\pi G \rho_{\ast} \left[\frac{a \left(1-e^2\right)}{1 + e \cos f} \sin \left(f + \Tilde{\omega}\right)\sin \Tilde{\imath}\right]^2}_{-\mathcal{R}}
\label{eq_H},
\end{equation}
\noindent where $\mu_{0}$ is the mass parameter of the solar system barycentre, $f$ is the true anomaly of the comet, $a$, $e$, $\Tilde{\imath}$, and $\Tilde{\omega}$ are the semimajor axis, eccentricity, orbital inclination, and argument of periapsis, respectively, in the solar system barycentric Galactic frame, and $\mathcal{R}$ is the disturbing function due to the Galactic tide, whose influence parallel to the Galactic plane is neglected \citep{1986Icar...65...13H}. The Keplerian orbital elements in the Galactic reference frame were converted from the counterparts in the ecliptic reference frame (see Appendix \ref{app_con_orb}). Since our focus was on secular variation in the orbit, following  \citet{1986Icar...65...13H}, we averaged the Hamiltonian in Equation (\ref{eq_H}) over the orbital period, thereby obtaining the orbit-averaged disturbing function
\begin{equation}
\bar{\mathcal{R}} = -\pi G \rho_{\ast} a^2 \left(1 - e^2 + 5e^2 \sin^{2} \Tilde{\omega} \right) \sin^{2} \Tilde{\imath}
\label{eq_R}.
\end{equation}
Given the relation between the periapsis distance $q$, the semimajor axis $a$, and the eccentricity $e$ of the orbit, namely, $q = a\left(1-e\right)$, we can write the time derivative of the periapsis distance as
\begin{equation}
\dot{q} = \left(1-e\right) \dot{a} - a \dot{e}
\label{eq_qdot},
\end{equation}
\noindent in which $\dot{a}$ and $\dot{e}$ can be computed by applying Lagrange's planetary equations \citep[e.g.,][]{2005ormo.book.....R}:
\begin{align}
\dot{a} & = \left(\frac{2}{n a}\right) \frac{\partial \bar{\mathcal{R}}}{\partial \chi}
\label{eq_adot}, \\
\dot{e} & = \frac{1}{n a^2 e} \left[\left(1-e^2\right)\frac{\partial \bar{\mathcal{R}}}{\partial \chi} - \sqrt{1 - e^2} \frac{\partial \bar{\mathcal{R}}}{\partial \Tilde{\omega}}\right]
\label{eq_edot}.
\end{align}
\noindent Here, $n$ is the mean motion and $\chi = -n t_{\rm p}$, in which $t_{\rm p}$ is the time of periapsis passage. Substituting with Equation (\ref{eq_R}), we can derive
\begin{align}
\dot{a} & = 0
\label{eq_adot2}, \\
\dot{e} & = 5 \pi G \rho_{\ast} \frac{e \sqrt{1-e^2}}{n}\sin^2 \Tilde{\imath} \sin 2\Tilde{\omega}
\label{eq_edot2}.
\end{align}
\noindent The first equation reveals that the semimajor axis of the orbit is a constant, as long as the perturbation is solely from the Galactic tide. Assuming there is no significant change in $\Tilde{\imath}$ and $\Tilde{\omega}$ with time, we can then apply the method of separation of variables to integrate the second equation, yielding
\begin{equation}
{\rm sech}^{-1} e_{\rm o} - {\rm sech}^{-1} e \approx 5 \pi G \rho_{\ast} \frac{\Delta t}{n} \sin^2 \Tilde{\imath} \sin 2 \Tilde{\omega}
\label{eq_De}
\end{equation}
\noindent in a time interval from $t_{\rm o}$ to $t_{\rm o} + \Delta t$. The eccentricity at initial time is denoted with the subscript ``o''.  Setting $\left|\Delta t \right| = 2 \pi / n$, namely, the orbital period of the comet, we can find the secular change in the periapsis distance between two successive returns of the comet due to the Galactic tide to be
\begin{align}
\nonumber
\Delta q & = q - q_{0} \\
& \approx a \left[e_{\rm o} - {\rm sech} \left({\rm sech}^{-1} e_{\rm o} \pm 10\pi^2 \frac{G \rho_{\ast}}{\mu_{0}} a^3 \sin^2 \Tilde{\imath} \sin 2 \Tilde{\omega} \right) \right]
\label{eq_Dq}.
\end{align}
\noindent Here, the mean motion has been substituted with the relation $\mu_{0} = n^2 a^3$, the plus sign is taken if the integration in Equation (\ref{eq_De}) is backward in time, and the minus sign is taken otherwise. Substitution with the obtained values of pertinent orbital elements of the original barycentric orbit referenced to the Galactic plane, we attained the periapsis distance of the previous return of the comet to be $88 \pm 6$ au. Therefore, our analytical result strongly favours that C/2019 E3 is a dynamically new comet from the Oort spike. In other words, the current return of the comet is most likely its first ever passage into the planetary region since it was ejected to the outer edge of the solar system. Given its long orbital period ($\sim\!6.6$ Myr), it is highly unlikely that the observed activity of the comet would be caused by retained heat from the previous return. 

In a similar fashion, we investigated the next return of the comet by integrating the 5,001 clones forwards in time until they reached a postperihelion heliocentric distance of $r_{\rm H} = 250$ au using completely the same force model in {\tt mercury6}. We append the statistics of the orbital elements of the ``future" orbit to Table \ref{tab:orb_OF}, from which we can notice that, if the comet survives the forthcoming perihelion passage, its orbital energy will decrease as a consequence of planetary perturbations. The change in the orbital energy is within the expected range of other large-perihelion comets from the Oort spike \citep{2017MNRAS.472.4634K}. In the same manner yet with the Keplerian orbital elements of the future barycentric orbit, we obtained the change in the periapsis distance between the current and next returns is $\Delta q = -3.19 \pm 0.07$ au, corresponding to an even smaller periapsis distance of $7.11 \pm 0.07$ au for the next return of the comet.

To check the reliability of our results from the analytical approach, we still employed {\tt mercury6} and integrated the nominal orbit of the comet (Table \ref{tab:orb}) both backwards and forwards in time until the previous and future returns were reached. We reduced the list of massive bodies to the eight major planets, Pluto, and the three most massive asteroids, so as to reduce the computational cost as much as possible. We assumed the linear theory for uncertainty propagation, whereby the error in the periapsis distance can be evaluated from
\begin{equation}
\sigma_{q} \approx \left[\left(\frac{\partial q}{\partial {\bf E}_{\rm o}} \right) {\bf C}_{\rm o} \left(\frac{\partial q}{\partial {\bf E}_{\rm o}} \right)^\top \right]^{1/2}
\label{eq_q_err}.
\end{equation}
\noindent Here, ${\bf E}_{\rm o}$ and ${\bf C}_{\rm o}$ are, respectively, the Keplerian orbital elements and the corresponding covariance matrix at initial time. We computed the partial derivatives using finite differences numerically in {\tt mercury6}, thus obtaining the previous and future periapsis distances of the comet to be $q = 72 \pm 4$ au and $7.59 \pm 0.06$ au, respectively. The result for the previous return is in reasonable agreement with the counterpart from the analytical approach at the $2\sigma$ level, given the approximations in the analytical approach. Yet the difference between the results for the next perihelion is at the $5\sigma$ level. Such a worse discrepancy is not surprising whatsoever, in that the comet will not be only perturbed by the Galactic tide, but also more profoundly by the major planets when it is well within the planetary region, which we completely ignored in the analytical approach. Despite these, the predicted trends for the periapsis distance of the comet, which is expected to further decrease in the next return, are consistent in both approaches.

We are fully aware of two major drawbacks in our analysis for the orbital evolution that nongravitational effects and gravitational perturbations from nearby passing stars were not taken into account. To address the former, we still utilised {\tt Find\_Orb} yet assumed a nongravitational acceleration stemming from sublimation of supervolatiles scaled as $r_{\rm H}^{-2}$ and included the radial, transverse, and normal nongravitational parameters defined by \citet{1973AJ.....78..211M} as additional free parameters to fit the astrometric observations of the comet. The result was that including the nongravitational parameters brought forth no clear improvement in the astrometric residuals of the solution, and that values of the best-fit nongravitational parameters are all statistically consistent with zero (signal-to-noise ratios $\sim\!0.1$). We therefore conclude that nongravitational effects of C/2019 E3 are negligible.

Now we proceed to discuss the second drawback. Using the Gaia DR2 catalogue and accounting for incompleteness, \citet{2018A&A...616A..37B} inferred a nontrivial stellar encounter rate within 1 pc of the present-day solar system to be $\sim\!20 \pm 2$ Myr$^{-1}$. Indeed, as an example, strong perturbations on the orbital evolution of C/2014 UN$_{271}$ by stellar encounters were reported in \citet{2022A&A...660A.100D}. On the other hand, \citet{2017MNRAS.472.4634K} identified no stellar encounter sufficient to alter the dynamical status of the analysed comets. Given these, we prefer that our conclusion about C/2019 E3 being a dynamically new comet likely remains valid, despite of neglecting stellar perturbations. However, we do suggest that our estimates of the previous and next periapsis distances should be better regarded as preliminary results, which need to be robustly verified by means of N-body integration accounting perturbations from passing stars. The verification is beyond the scope of this work. We noticed a fairly recent update of the Catalogue of Cometary Orbits and their Dynamical Evolution\footnote{\url{https://pad2.astro.amu.edu.pl/comets/index.php}} (CODE) by \citet{2020A&A...640A..97K,2023arXiv231104063K} that 29 comets discovered between 2019 and early 2021, including C/2019 E3, were newly added to the database. Although their orbit determination for C/2019 E3 was based on available astrometric observations of a shorter arc spanning from 2015 January to 2023 May, they also arrived at the same conclusion as ours that C/2019 is a dynamically new comet, with a previous periapsis distance of $q = 60 \pm 5$ au in the model where only the Galactic tide was included, and $441 \pm 6$ au in the model where stellar perturbations were also taken into account. Unfortunately, the future orbital evolution of the comet is not available from the CODE catalogue. Nevertheless, the dynamical status of C/2019 E3 being a dynamically new comet appears to be conclusive.

\subsection{Comparison of the Known Ultra-distant Comets}
\label{ss_comp}

\begin{figure*}
\epsscale{1.2}
\begin{center}
\plotone{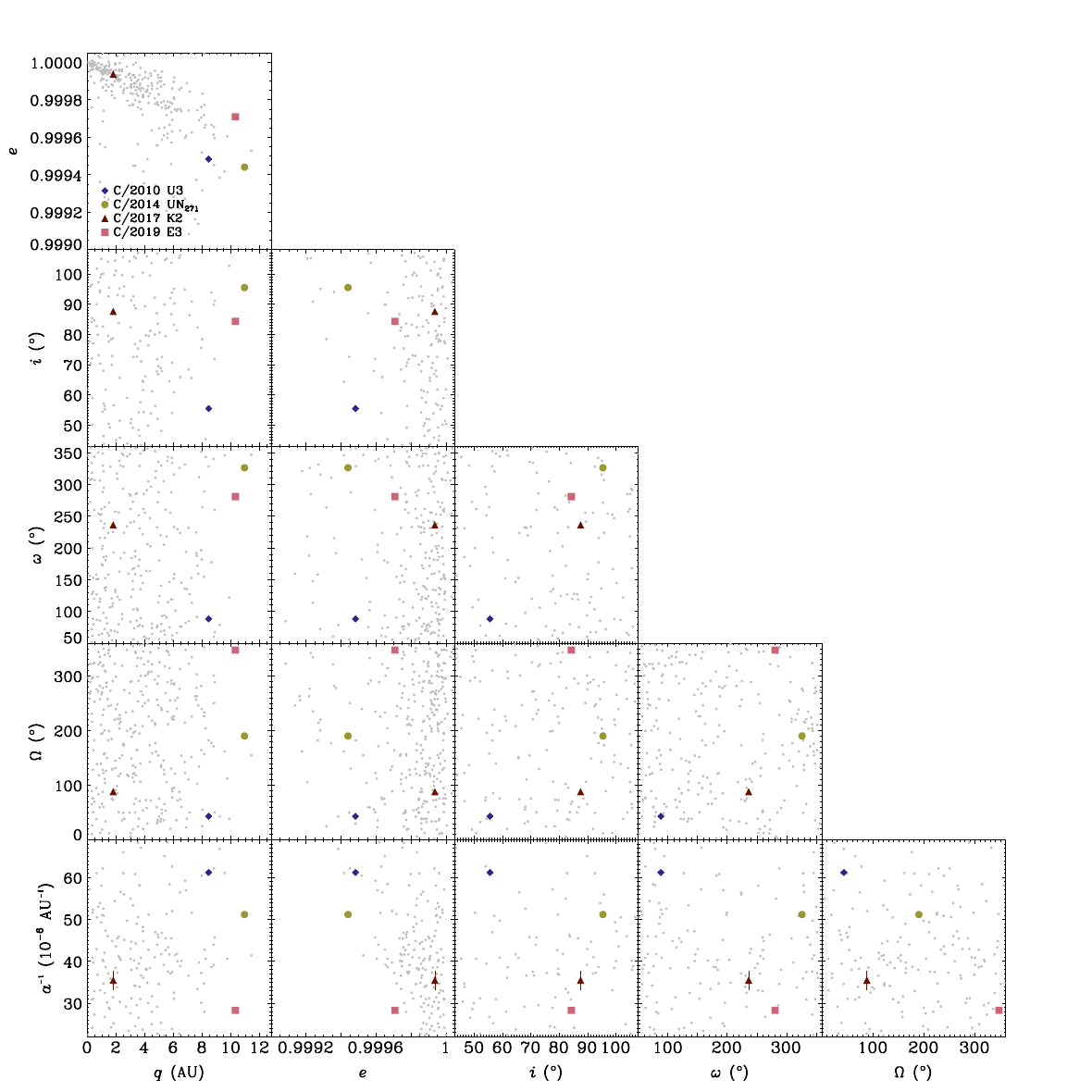}
\caption{
Comparison between periapsis distance ($q$), eccentricity ($e$), inclination ($i$), argument of periapsis ($\omega$), longitude of ascending node ($\Omega$), and reciprocal of the semimajor axis ($a^{-1}$) of original solar system barycentric J2000 ecliptic orbits of the four known ultra-distant comets C/2010 U3, C/2014 UN$_{271}$, C/2017 K2, and C/2019 E3 (colour coded by bold symbols, see the legend in the upper left). Background dots in grey are other long-period comets, whose original orbits together with those of the three previously known ultra-distant comets are based on the preferred solutions in the CODE catalogue. Associated $1\sigma$ formal errors in the orbital elements of the ultra-distant comets are also included in the plots. However, as they are in general much smaller than the displayed ranges here, they are basically invisible, except for C/2017 K2 in terms of its reciprocal of the semimajor axis.
\label{fig:orb_UDC}
} 
\end{center} 
\end{figure*}

Here we compare ultra-distant comets that exhibited preperihelion cometary activity at heliocentric distances $r_{\rm H} \ga 20$ au. Noteworthily, it came to our attention that \citet{2023A&A...678A.113K} included three astrometric measurements of comet C/2006 S3 (LONEOS) at a preperihelion heliocentric distance of $r_{\rm H} = 26.1$ au all from 1999 October 13 in their orbit determination, implying the comet as an additional ultra-distant comet. On the other hand, these astrometric measurements were found to be outliers with great residuals that could not fit any of the orbital solutions for the comet by these authors. After querying astrometry for the comet in the Minor Planet Center Database Search, we found that the three astrometric measurements no longer existed. Moreover, the earliest astrometry of the comet used in orbit determination by the Minor Planet Center and JPL Horizons was no earlier than 2006 August, when the comet was at $\sim\!14$ au from the Sun. As such, we suspect that the single-night observation of the comet from 1999 is erroneous and do not count C/2006 S3 as a known ultra-distant comet. It is also worth noting that comet C/1995 O1 (Hale-Bopp) was observed to exhibit activity at great heliocentric distances $\ga\!20$ au as well \citep{2008ApJ...677L.121S,2014Icar..236..136K}. However, the observed activity at $r_{\rm H} \ga 20$ au was all postperihelion, and there is no reported observation showing its activity at similarly great heliocentric distances preperihelion. Thus, it is not counted here either. As a result, there are four known samples of ultra-distant comets to date: C/2010 U3, C/2014 UN$_{271}$, C/2017 K2, and C/2019 E3.

First of all, let us compare the original orbits of the four ultra-distant comets. For C/2010 U3, C/2014 UN$_{271}$, and C/2017 K2, we simply adopt the ``preferred" solutions from the CODE catalogue. We show the comparison together with other long-period comets from the CODE catalogue in multidimensional space of the orbital elements in Figure \ref{fig:orb_UDC}, where we can find no obvious trend in the orbital distribution, except that they all have considerable orbital inclinations with respect to the ecliptic. In terms of original orbits, the known ultra-distant comets are not distinguishable from other long-period comets. 

Now we discuss the dynamical statuses of the known ultra-distant comets. According to the CODE catalogue, C/2010 U3 is confidently labelled as a dynamically new comet, whereas C/2017 K2 was formerly reported to be ``almost certainly" a dynamically old comet by \citet{2018A&A...615A.170K}. However, even though the observing arc of the comet was extended, its dynamical status turned out to be far more equivocal than previously expected, primarily owing to great uncertainties in a potential close encounter with a nearby passing star \citep{2022A&A...660A.100D}. As for C/2014 UN$_{271}$, \citet{2021ApJ...921L..37B} concluded the comet to be dynamically new, which is largely supported by \citet{2022A&A...660A.100D}, with the exception that the latter authors also identified a nontrivial number of cases where the comet entered the planetary region in the previous perihelion return. Together with the results for C/2019 E3 (Section \ref{ss_orbev}), there seems to be a fair chance that all of the four known ultra-distant comets are dynamically new. Yet this remains to be affirmed by force models with more precise kinematics of both the comets and nearby passing stars.

Next, we compare the physical properties of the four ultra-distant comets. Their general surface brightness profiles were all measured to be consistent with steady state, suggestive of prolonged sublimation of supervolatiles \citep{2017ApJ...847L..19J,2019AJ....157...65J,2019AJ....157..162H,2022ApJ...929L..12H,2022ApJ...933L..44K}. Morphologically, C/2019 E3 and C/2017 K2 highly resembled each other, as no obvious tail was visible at heliocentric distances $r_{\rm H} \ga 10$ au, implying the optical dominance of large-sized (submillimetre-scaled or greater) dust particles ejected at speeds no greater than a few metres per second. Conversely, at similar distances from the Sun, C/2010 U3 and C/2014 UN$_{271}$ showed obvious tails. In particular in the case of C/2010 U3, the Monte Carlo dust modelling by \citet{2019AJ....157..162H} suggested that the coma and tail of the comet consisted of much smaller dust grains of $\sim\!10$~\micron~ejected at speeds of $\la\!50$ m s$^{-1}$ and subjected to the Lorentz force in addition to the solar gravitation and radiation pressure. For C/2014 UN$_{271}$, we are unaware of any Monte Carlo dust modelling in the published literature. Yet the syndyne analysis by \citet{2021PSJ.....2..236F} suggests that the observed dust coma and tail comprised of dust grains of submillimetre scale ejected in the sunward hemisphere of the nucleus at faster speeds of $\sim\!10$ m s$^{-1}$. Thus, although the known ultra-distant comets were likely all driven by sublimation of supervolatiles in steady state, there possibly exists a diversity in the physical properties of their dust environments. Nevertheless, the discoveries of ultra-distant comets still act as a serious challenge to the classical activity model for comets, which predicts no cometary activity whatsoever at heliocentric distances $\ga\!10$ au due to interparticle cohesion being overlarge for the drag force from outgassing supervolatiles to overcome \citep{2015A&A...583A..12G,2019AJ....157...65J}. Recent promising resolutions of the dilemma include accounting gas diffusion inside pebbles of which a cometary nucleus is comprised \citep{2020A&A...636L...3F} and sublimation through a porous mantle \citep{2022ApJ...924...37B}, but they remain to be further verified with more observations of ultra-distant comets.

Before concluding this section, we also compare the activity of the four ultra-distant comets. At heliocentric distances $\ga\!20$ au, the apparent magnitudes of these comets were not greatly different from each other, with C/2014 UN$_{271}$ being somewhat brighter, presumably at least in part attributed to its exceptionally enormous nucleus size \citep{2022A&A...659L...1L,2022ApJ...929L..12H}. However, as these comets neared the Sun, there seem to be two primary types to which the four ultra-distant comets belong. The first type includes C/2017 K2 and C/2019 E3, whose activity in general monotonically increased in highly similar manners without observational evidence of outbursts. The other type contains C/2010 U3 and C/2014 UN$_{271}$, which exhibited clear outbursts in brightness \citep{2019AJ....157..162H,2022ApJ...933L..44K}. While we cannot completely rule out the possibility of outbursts of C/2017 K2 and C/2019 E3 evading the available observations, it still seems reasonable to conclude that outbursts at comets of the first type occurred on a rarer basis if compared to the counterparts of the other type.

At the current stage, as we are only beginning to discover ultra-distant comets and there are only four known samples, we know practically little about this population itself and have very few clues about how they can be active even at great distances from the Sun. In order to acquire a better understanding of these comets, far more members of this population are desired. With the advent of state-of-the-art all-sky surveys such as the Legacy Survey of Space and Time programme to be conducted starting from 2025 at the Vera C. Rubin Observatory \citep{2019ApJ...873..111I}, it is suggested that hundreds to thousands of long-period comets with perihelion distances $\ga\!5$ au will be discovered in the near future \citep{2016AJ....152..103S}. These new observations would allow for an advancement in our understanding of the ultra-distant comet population and their activity.


\section{Summary}
\label{sec_sum}

In this paper, we presented analyses of serendipitous archival observations of comet C/2019 E3. Key findings of our study include:
\begin{enumerate}
\item We identified C/2019 E3 (ATLAS) as the fourth known ultra-distant comet that displayed prolonged activity at heliocentric distances $r_{\rm H} \ga 20$ au preperihelion, after C/2010 U3 (Boattini), C/2014 UN$_{271}$ (Bernardinelli-Bernstein), and C/2017 K2 (PANSTARRS).
\item The measurements of the surface brightness profile of the comet reveal that the activity was consistent with steady state, thus suggesting the mass loss being driven by sublimation of supervolatiles (e.g., CO, CO$_{2}$), given the great heliocentric distances of the comet. The effective scattering cross-section of dust increased steadily as the comet approached the Sun, varying with heliocentric distance as $r_{\rm H}^{-1.5 \pm 0.4}$ over the monitored course from 2012 to 2019. We estimated a total and net mass-loss rate of dust of $\ga\!10^{2}$ kg s$^{-1}$ and $\ga\!10$ kg s$^{-1}$, respectively, for the comet.
\item Our photometry of the comet indicates its colour was similar to those of other long-period comets, except in $g-r$, which was measured to be slightly bluer. Based on our photometry, assuming a geometric albedo of $p_r = 0.05$, and adopting several empirical thermophysical models for sublimation of CO and CO$_{2}$, we constrained the nucleus of the comet to be greater than $\sim\!3$ km in diameter.
\item The apparent morphology of the comet basically remained circularly symmetric without an obvious tail. Results from our Monte Carlo dust modelling suggest that the observed dust environment of the comet was most likely optically dominated by mm-scaled dust grains ejected from the sunlit hemisphere of the nucleus at speeds $\la\!0.4$ m s$^{-1}$.
\item We performed an improved orbit determination for the comet with our astrometry, based upon which we analytically derived that the comet is likely dynamically new and that its periapsis distance will further shrink in the next return due to the Galactic tide, assuming the comet will physically survive. Although the detailed values of the periapsis distances in the previous and next returns are preliminary and should be verified by a more sophisticated dynamical model where perturbations from stellar encounters are taken into account, the dynamical status of C/2019 E3 being a dynamically new comet is incontestable.
\item Our comparison of the four known ultra-distant comets reveals no obvious trend seen in their orbital elements. Nor do they exhibit outstanding orbital traits if compared to other long-period comets. While there possibly exists a diversity in the physical properties of their dust environments, there seems to be two types into which these comets can be grouped -- 1) C/2017 K2 and C/2019 E3 increased their activity basically in a monotonic manner as they approached the Sun, and 2) C/2010 U3 and C/2014 UN$_{271}$ exhibited obvious outbursts in brightness at similar heliocentric distances. We need more data of this distant comet population before we can achieve a better understanding of them and their activity.
\end{enumerate}

\section*{acknowledgements}
We thank Davide Farnocchia, Shoucun Hu, and Paul Wiegert for insightful discussions, Maria Womack and an anonymous referee for their valuable comments and suggestions on our manuscript, as well as observers who submitted meticulous astrometric measurements of C/2019 E3 to the Minor Planet Center. 

This project used data obtained with the Dark Energy Camera (DECam), which was constructed by the Dark Energy Survey (DES) collaboration. Funding for the DES Projects has been provided by the US Department of Energy, the US National Science Foundation, the Ministry of Science and Education of Spain, the Science and Technology Facilities Council of the United Kingdom, the Higher Education Funding Council for England, the National Center for Supercomputing Applications at the University of Illinois at Urbana-Champaign, the Kavli Institute for Cosmological Physics at the University of Chicago, Center for Cosmology and Astro-Particle Physics at the Ohio State University, the Mitchell Institute for Fundamental Physics and Astronomy at Texas A\&M University, Financiadora de Estudos e Projetos, Funda\c{c}\~{a}o Carlos Chagas Filho de Amparo {\`a} Pesquisa do Estado do Rio de Janeiro, Conselho Nacional de Desenvolvimento Cient{\'i}fico e Tecnol{\'o}gico and the Minist{\'e}rio da Ci\^{e}ncia, Tecnologia e Inova\c{c}\~{a}o, the Deutsche Forschungsgemeinschaft and the Collaborating Institutions in the Dark Energy Survey.

The Collaborating Institutions are Argonne National Laboratory, the University of California at Santa Cruz, the University of Cambridge, Centro de Investigaciones En{\'e}rgeticas, Medioambientales y Tecnol{\'o}gicas-Madrid, the University of Chicago, University College London, the DES-Brazil Consortium, the University of Edinburgh, the Eidgen{\"o}ssische Technische Hochschule (ETH) Z{\"u}rich, Fermi National Accelerator Laboratory, the University of Illinois at Urbana-Champaign, the Institut de Ci{\`e}ncies de l'Espai (IEEC/CSIC), the Institut de F{\'i}sica d'Altes Energies, Lawrence Berkeley National Laboratory, the Ludwig-Maximilians Universit{\"a}t M{\"u}nchen and the associated Excellence Cluster Universe, the University of Michigan, NSF’s NOIRLab, the University of Nottingham, the Ohio State University, the OzDES Membership Consortium, the University of Pennsylvania, the University of Portsmouth, SLAC National Accelerator Laboratory, Stanford University, the University of Sussex, and Texas A\&M University.

Based on observations at Cerro Tololo Inter-American Observatory, NSF’s NOIRLab (NOIRLab Prop. IDs 2013A-0741 by PI D. Schlegel, 2014A-0620 by A. Casey, 2014A-0270 by C. Grillmair, and 2019A-0305 by A. Drlica-Wagner), which is managed by the Association of Universities for Research in Astronomy (AURA) under a cooperative agreement with the National Science Foundation.

Pan-STARRS is supported by the National Aeronautics and Space Administration under Grants 80NSSC18K0971 and 80NSSC21K1572 issued through the SSO Near Earth Object Observations Program.

The national facility capability for SkyMapper has been funded through ARC LIEF grant LE130100104 from the Australian Research Council, awarded to the University of Sydney, the Australian National University, Swinburne University of Technology, the University of Queensland, the University of Western Australia, the University of Melbourne, Curtin University of Technology, Monash University and the Australian Astronomical Observatory. SkyMapper is owned and operated by The Australian National University's Research School of Astronomy and Astrophysics.

Based on Very Large Telescope (VLT) Survey Telescope (VST)
OmegaCAM data obtained from the ESO Science Archive Facility under request No. 864155.

Based on observations obtained with the Samuel Oschin Telescope 48-inch and the 60-inch Telescope at the Palomar Observatory as part of the Zwicky Transient Facility project. ZTF is supported by the National Science Foundation under Grant No. AST-2034437 and a collaboration including Caltech, IPAC, the Weizmann Institute for Science, the Oskar Klein Center at Stockholm University, the University of Maryland, Deutsches Elektronen-Synchrotron and Humboldt University, the TANGO Consortium of Taiwan, the University of Wisconsin at Milwaukee, Trinity College Dublin, Lawrence Livermore National Laboratories, and IN2P3, France. Operations are conducted by COO, IPAC, and UW.

This research has made use of data and/or services provided by the International Astronomical Union's Minor Planet Center and the facilities of the Canadian Astronomy Data Centre operated by the National Research Council of Canada with the support of the Canadian Space Agency. The work was supported by Science and Technology Development Fund, Macau SAR, through grant No. 0016/2022/A1 to M.T.H.


\vspace{5mm}
\facilities{Blanco (DECam), PO: 1.2m (ZTF), PS1, SkyMapper, VST (OmegaCAM)}

\software{{\tt Find\_Orb}, {\tt IDL}, {\tt mercury6} \citep{1999MNRAS.304..793C}}

\clearpage
\appendix

\section{Equation of Motion Perturbed by Galactic Tide}
\label{app_eom}

Adopting the Galactic tide model by \citet{1986Icar...65...13H}, \citet{2005CeMDA..93..229F} presented the equations of motion of a comet perturbed by the Galactic tide in a mixture of fixed and rotating Cartesian Galactic coordinates. Starting with their results, we express the equation of motion of the comet in the J2000 barycentric Cartesian Galactic coordinates $\left(x,y,z \right)^\top$ in a matrix form as
\begin{equation}
\begin{pmatrix}
\Ddot{x}\\
\Ddot{y}\\
\Ddot{z}
\end{pmatrix}
= -\frac{\mu_{0}}{r^3}
\begin{pmatrix}
x\\
y\\
z
\end{pmatrix}
- 
\begin{pmatrix}
\mathcal{G}_1 \cos 2\Omega_{\odot} t & \mathcal{G}_1 \sin 2\Omega_{\odot} t & 0 \\
\mathcal{G}_1 \sin 2\Omega_{\odot} t & -\mathcal{G}_1 \cos 2\Omega_{\odot} t & 0 \\
0 & 0 & \mathcal{G}_3
\end{pmatrix}
\begin{pmatrix}
x\\
y\\
z
\end{pmatrix}
\label{eq_eom},
\end{equation}
\noindent where $r = \sqrt{x^2 + y^2 + z^2}$ is the barycentric distance of the comet, $\Omega_{\odot} = -26$ km s$^{-1}$ kpc$^{-1}$ is the angular speed of the Sun around the Galactic centre, $\mathcal{G}_{1} = -\Omega_{\odot}^2$ and $\mathcal{G}_{3} = 4\pi G \rho_{\ast}$ are the Oort constants, and $t$ is time from J2000. The first and second terms on the right-hand side of Equation (\ref{eq_eom}) correspond to the contribution from the gravity of the solar system barycentre and that from the Galactic tide, respectively. 

Let ${\bf R}$ denote the $3 \times 3$ transformation matrix from the ecliptic reference frame $\left(X,Y,Z \right)^\top$ to the galactic one $\left(x,y,z\right)^\top$ at epoch J2000, when the obliquity of the ecliptic is $\epsilon = 23\degr26\arcmin 21\farcs448$, the R.A. and decl. coordinates of the north Galactic pole are $\alpha_{\rm G} = 192\fdg85948$ and $\delta_{\rm G} = +27\fdg12825$, respectively, and the Galactic longitude of the ascending node of the Galactic plane on the celestial equator is $l_{\Omega} = 32\fdg93192$ \citep{1997ESASP1200.....E}. As such, the transformation from the ecliptic reference frame to the Galactic one can be achieved by rotating first about the first axis by $\epsilon$ clockwise, then counterclockwise around the third axis by $\pi/2 + \alpha_{\rm G}$, followed by counterclockwise around the first axis by $\pi/2 - \delta_{\rm G}$, and finally clockwise about the third axis by $l_{\Omega}$, i.e.,
\begin{equation}
{\bf R} = {\bf R}_{3} \left(-l_{\Omega}\right) {\bf R}_{1} \left(\frac{\pi}{2} - \delta_{\rm G}\right) {\bf R}_{3} \left(\frac{\pi}{2} + \alpha_{\rm G}\right) {\bf R}_1 \left(-\epsilon \right)
\label{eq_rotmat_EC2GAL}.
\end{equation}
\noindent Here, ${\bf R}_{j} \left(\theta\right)$ ($j=1,2,3$) represents the $3\times3$ rotation matrix that performs a rotation about the $i$-th axis of the coordinate system by an angle of $\theta$ ($>0$ for counterclockwise rotation, otherwise clockwise) in $\mathbb{R}^3$. Our result for the transformation matrix to ten decimal digits is:
\begin{equation}
{\bf R} = 
\begin{pmatrix}
-0.0548755604 & -0.9938213791 & -0.0964766261 \\
+0.4941094279 & -0.1109907334 & +0.8622858751 \\
-0.8676661490 & -0.0003515899 & +0.4971471917
\end{pmatrix}
\label{eq_R_val}.
\end{equation}

Thereby, the acceleration of the comet due to the Galactic tide in the Cartesian ecliptic coordinates at J2000 can be easily obtained from the second term on the right-hand side of Equation (\ref{eq_eom}) with the transformation matrix as
\begin{equation}
\begin{pmatrix}
\Ddot{X} \\
\Ddot{Y} \\
\Ddot{Z}
\end{pmatrix}_{\rm tide}
= -{\bf R}^{\top}
\begin{pmatrix}
\mathcal{G}_1 \cos 2\Omega_{\odot} t & \mathcal{G}_1 \sin 2\Omega_{\odot} t & 0 \\
\mathcal{G}_1 \sin 2\Omega_{\odot} t & -\mathcal{G}_1 \cos 2\Omega_{\odot} t & 0 \\
0 & 0 & \mathcal{G}_3
\end{pmatrix}
{\bf R}
\begin{pmatrix}
X\\
Y\\
Z
\end{pmatrix}
\label{eq_a_tide},
\end{equation}
\noindent which we implemented and added to the subroutine {\tt mfo\_user}() in {\tt mercury6}.

\section{Conversions Between Ecliptic \& Galactic Orbital Elements}
\label{app_con_orb}

The transformation between the Galactic and ecliptic reference frames is purely rotational and therefore lengths are conserved; the only influenced Keplerian orbital elements are the inclination $i$, longitude of ascending node $\Omega$, and argument of periapsis $\omega$. In the following, we present the conversions from the ecliptic reference frame to the Galactic one for the three aforementioned Keplerian orbital elements.
\begin{align}
\cos \Tilde{\imath} = & \left(R_{31} \sin \Omega - R_{32} \cos \Omega \right) \sin i + R_{33} \cos i
\label{eq_cos_I}, \\
\sin \Tilde{\Omega} = &
\frac{\left(R_{11} \sin \Omega - R_{12} \cos \Omega \right) \sin i + R_{13} \cos i}{\sin \Tilde{\imath}}
\label{eq_Om_sin},\\
\cos \Tilde{\Omega} = &
-\frac{\left(R_{21} \sin \Omega - R_{22} \cos \Omega \right) \sin i + R_{23} \cos i}{\sin \Tilde{\imath}}
\label{eq_Om_cos},\\
\sin \Tilde{\omega}' 
= & \frac{\left(R_{31} \cos \Omega + R_{32} \sin \Omega \right) \cos \omega' - \left[\left(R_{31} \sin \Omega - R_{32}\cos \Omega\right) \cos i - R_{33} \sin i\right]\sin \omega'}{\sin \Tilde{\imath}}
\label{eq_o_sin},\\
\begin{split}
\cos \Tilde{\omega}' = &
\frac{R_{11} \left[R_{22} \sin i \cos \omega' + R_{23} \left(\sin \Omega \sin \omega' - \cos i \cos \Omega \cos \omega' \right) \right]}{\sin \Tilde{\imath}} - 
\\ &
\frac{R_{12} \left[R_{21} \sin i \cos \omega' + R_{23} \left(\cos \Omega \sin \omega' + \cos i \sin \Omega \cos \omega' \right) \right]}{\sin \Tilde{\imath}} - 
\\ &
\frac{R_{13} \left[R_{21} (\sin \Omega \sin \omega' - \cos i \cos \Omega \cos \omega') - R_{22} \left(\cos \Omega \sin \omega' + \cos i \sin \Omega \cos \omega' \right) \right]}{\sin \Tilde{\imath}}
\end{split}
\label{eq_o_cos}.
\end{align}
\noindent In the above equations, $R_{kl}$ denotes the element in $k$-th row and $l$-th column of transformation matrix ${\bf R}$, the symbol $\Tilde{\square}$ denotes the corresponding Keplerian orbital elements in the Galactic reference frame, $\square' = \square + f$, where $f$ is the true anomaly, a quantity calculatable from the mean anomaly and independent from the selection of the reference system.

The conversions from the Galactic reference frame to the ecliptic one for the three Keplerian orbital elements can be conveniently obtained by simply swapping the Galactic and ecliptic orbital elements as well as the indices of rows and columns in Equations (\ref{eq_cos_I})-(\ref{eq_o_cos}), thanks to the property of ${\bf R}^{-1} = {\bf R}^\top$.


\end{document}